\documentstyle[prd,aps]{revtex}

\begin{document}

\draft

\title{
Scalar sea contributions\\
to spin 1/2 baryon structure\\
and magnetic moments
}

\author{
V.~Gupta\cite{email1},\ 
R.~Huerta\cite{email2},\ 
and
G.~S\'anchez-Col\'on\cite{email3}.
}
\address{
Departamento de F\'{\i}sica Aplicada. \\
Centro de Investigaci\'on y de Estudios Avanzados del IPN.\\
Unidad M\'erida. \\ 
A.P. 73, Cordemex 97310, M\'erida, Yucat\'an, MEXICO. \\
}

\date{February 23, 1996}

\maketitle

\begin{abstract}
We treat the baryon as a composite system made out of a \lq\lq core" of three
quarks (as in the standard quark model) surrounded by a \lq\lq sea" (of gluons
and $q\bar{q}$-pairs) which is specified by its total quantum numbers like
flavor, spin and color.
Specifically, we assume the sea to be a flavor octet with spin 0 but no color.
The general wavefunction for spin 1/2 baryons with such a sea component is
given and an application to the magnetic moments is considered.
Numerical analysis shows that the scalar sea can provide a good fit to the
magnetic moment data {\em using experimental errors}.
\end{abstract}


\section{Introduction}

Attempts to understand the static properties of hadrons in the framework of
the standard quark model (SQM) have had limited success.
The belief that valence quarks were responsible for the spin of the proton has
been shattered by recent experiments\cite{1}.

The naive valence quark picture of hadron structure is a simplification which
does not properly take into account the fact that quarks interact through
color forces mediated by vector gluons.
The existence of the quark-gluon interaction, in QCD, implies that a hadron
should be viewed as consisting of valence quarks surrounded by a \lq\lq sea"
which contains gluons and virtual quark-antiquark ($q\bar{q}$) pairs.
Deep inelastic lepton-nucleon scattering has shown the existence of a sea
component and its importance for nucleon structure functions.
It is thus necessary to understand how this sea contributes to the baryon spin
and other low energy properties.

Several authors\cite{2,3,4,5,6} have studied the effect of the sea
contributions on the hadron structure and  the static properties of baryons.
Some consider the sea as a single gluon or a $q\bar{q}$-pair.
However the sea, in general, consists of any number of gluons and
$q\bar{q}$-pairs.
In this paper we \lq\lq model" the general sea by its total quantum numbers
(flavor, spin and color) which are such that the sea wave function when
combined with the valence quark wave function gives the desired quantum
numbers for the physical hadron.
In particular, we explore the consequences of a \lq\lq sea" with flavor and
spin but no color\cite{7} for the low energy properties of the spin 1/2 baryon
octet $(p,n,\Lambda,\dots)$.

For simplicity, we consider a flavor octet sea with spin 0.
We find that the scalar (spin 0) sea described by two parameters gives
a very good fit to the magnetic moment data using actual experimental errors
which is our primary motivation.

In the next section, we discuss the wavefunctions for the physical baryons
constructed from the SQM and from our model sea.
Then, we obtain the magnetic moments from the modified
wavefunction and give a general discussion of the results.
The last section gives a summary and discussion. 

\section{Spin 1/2 octet baryon wavefunctions with sea}

For the lowest-lying baryons, in the SQM, the three valence quarks are taken
to be in a relative $S$-wave states so that the space wavefunction is symmetric
and the color wavefunction is totally antisymmetric to give a color singlet
baryon.
Consequently, the flavor-spin wavefunction is totally symmetric.
For the $SU(3)$ flavor octet spin 1/2 baryons we denote this SQM or $q^3$
wavefunction by $\tilde{B}({\bf 8},1/2)$, the argument 1/2 refers to spin.
The corresponding states are denoted by $\tilde{p}$, $\tilde{\Sigma}^{+}$, etc.

In our picture, the physical baryon octet $B({\bf 8}, 1/2)$ states are formed
out of the $q^{3}$ \lq\lq core" baryons described by
$\tilde{B}({\bf 8},1/2)$ surrounded by a sea consisting of $q\bar{q}$-pairs
and gluons.
This sea is assumed to be color singlet but has flavor and spin properties
such that its wavefunction $\Phi_s$ (subscript \lq\lq $s$" for sea) when
combined with the core baryons gives the physical baryons, that is

\begin{equation}
\tilde{B}({\bf 8},1/2)\otimes\Phi_{s} = B({\bf 8},1/2).
\label{uno}
\end{equation}

The possible SU(3) flavor and spin transformation properties the sea can have
are $\bf 1,8,10,\bar{10},27$ for flavor and 0 for spin.
The corresponding wavefunctions are denoted by $S(N)$ ($N={\bf 1,8,\dots}$)
and $H_{0}$.
So, in general $\Phi_{s}$ is a combination of $S(N)H_{0}$.
The $SU(3)$ symmetric and spinless (color singlet) sea implicit in the SQM is
discribed
by the wavefunction $S({\bf 1})H_{0}$.
Such a color singlet sea would require at least two gluons by themselves or a
$q\bar{q}$-pair.
The sea described by the wavefunctions $S({\bf 8})H_{0}$ would require a
minimum of one $q\bar{q}$-pair.
Gluons by themselves cannot give flavorful sea while, at least two
$q\bar{q}$-pairs are required to give a sea with $N={\bf 10,\bar{10},27}$. 
Guided by simplicity and the above remarks we limit our discussion to a sea
described by the wavefunctions $S({\bf 1})H_{0}$ and $S({\bf 8})H_{0}$.
The color singlet sea discribed by these wavefunctions can, in general, have
any number of $q\bar{q}$-pairs and gluons consistent with its total flavor and
spin quantum numbers.
We also refer to a spin 0 sea as a scalar sea.

The total flavor-spin wavefunction of a spin up ($\uparrow$) physical baryon
which consists of 3 valence quarks and a sea component (as discussed above)
can be written schematically as

\begin{equation}
B(1/2\uparrow)
=\tilde{B}({\bf 8},1/2\uparrow)H_{0}S({\bf 1})
+\sum_{N}a(N)\left[\tilde{B}({\bf 8},1/2\uparrow)H_{0}
\otimes S({\bf 8})\right]_{N}
\label{dos}
\end{equation}

\noindent
The normalization not indicated here is discussed later.
The first term is the usual $q^{3}$-wavefunction of the SQM (with a trivial
sea), in this term the sea is a flavor singlet.
The second term in Eq.~(\ref{dos}) contains a scalar sea which transforms as a
flavor octet.
The various $SU(3)$ flavor representations obtained from
$\tilde{B}({\bf 8})\otimes S({\bf 8})$ are labelled by
$N={\bf 1,8_{F},8_{D},10,\bar{10},27}$.
As it stands, Eq.~(\ref{dos}) represents a spin 1/2$\uparrow$ baryon which is
not {\em a pure flavor octet} but has an admixture of other $SU(3)$
representations weighted by the unspecified constants $a(N)$.
It will be a flavor octet if $a(N)=0$ for $N={\bf 1,10,\bar{10},27}$.
The color wavefunctions have not been indicated as the three valence quarks in
the core $\tilde{B}$ and the sea (by assumption) are in a color singlet state.

For our applications we adopt the phenomenological wavefunction given in
Eq.~(\ref{dos}), where the physical spin 1/2 baryons have admixtures of flavor
$SU(3)$ determined by the coefficients $a(N)$,
$N={\bf 1,10,\bar{10},27}$.
As we shall see, such a wavefunction which respects the isospin and
hypercharge properties of the usual spin 1/2 baryon states is general enough
to provide an excellent fit to the magnetic moments data.
Surprisingingly, only 2 or 3 of the six parameters in Eq.~(\ref{dos}) are
needed for this purpose.
For the moment we discuss the general wavefunction in Eq.~(\ref{dos}) as it is.
Incidentally, such a wavefunction could arise in general since we know flavor
is broken by mass terms in the QCD Lagrangian.

The sea isoespin multiplets contained in the octet $S({\bf 8})$ are denoted as

\begin{equation}
(S_{\pi^+},S_{\pi^0},S_{\pi^-}),\qquad
(S_{K^+},S_{K^0}),\qquad
(S_{\bar{K}^0},S_{K^-}),\quad
\hbox{and}\quad S_{\eta}.
\label{cuatro}
\end{equation}     
 
\noindent
The suffix on the components label the isospin and hypercharge quantum numbers.
For example, $S_{\pi^+}$ has $I=1$, $I_3=1$ and $Y=0$; $S_{K^-}$ has
$I=1/2$, $I_3=-1/2$, and $Y=-1$; etc.
These flavor quantum numbers when combined with those of the three valence
quarks states $\tilde{B}$ will give the observed $I$, $I_3$, and $Y$ for the
physical states $B$.
The flavor combinations in the second term in Eq.~(\ref{dos}) imply
that the physical states $B(Y,I,I_3)$ are expressed as a sum of products of
$\tilde{B}(Y,I,I_3)$ and the sea components $S(Y,I,I_3)$, weighted by
coefficients which are linear combinations of the coefficients $a(N)$.
Schematically, the flavor content of the second term in
Eq.~(\ref{dos}) is of the form (suppressing $I_3$)

\begin{equation}
B(Y,I)=
\sum_i \alpha_i(Y_1,Y_2,I_1,I_2)\left[\tilde{B}(Y_1,I_1)S(Y_2,I_2)\right]_i
\label{cinco}
\end{equation}

\noindent
where the sum is over all $Y_i$, $I_i$, ($i=1,2$); such that: $Y=Y_1+Y_2$ and 
$\bf I=I_1+I_2$. 
The flavor content of $B(Y,I,I_3)$ in terms of $\tilde{B}(Y,I,I_3)$ and sea
components are given in Table~\ref{tabla1}.
The corresponding coefficients $\bar{\beta}_i$, $\beta_i$, etc. expressed in
terms of the coefficients $a(N)$ are recorded in Table~\ref{tabla2}.
In Table~\ref{tabla1} we have denoted $\tilde{B}(Y,I,I_3)$ and $S(Y,I,I_3)$ by
appropiate symbols, e.g., $\tilde{B}(1,1,1/2)$ by $\tilde{p}$, $S(0,1,1)$ by
$S_{\pi^+}$, etc.
In Tables~\ref{tabla1} and \ref{tabla2} for the reduction of
$\tilde{B}({\bf 8})\otimes S({\bf 8})$ into various $SU(3)$ representations we
have followed the convention used by Carruthers~\cite{carruthers}.

The normalization of the physical baryons wavefunction in Eq.~(\ref{dos}) can
be obtained by using
$\langle \tilde{B}(Y,I,I_3)|\tilde{B}(Y',I',I'_3)\rangle =
\langle S(Y,I,I_3)|S(Y',I',I'_3)\rangle =
\delta_{YY'}\delta_{II'}\delta_{I_3I'_3}$.
However, it should be noted that the normalization are different, in general,
for each $B(Y,I)$ state.
This is because not all $a(N)$ contribute to a given
$(Y,I)$-multiplet as is clear from Tables~\ref{tabla1} and \ref{tabla2}.
For example, $a({\bf 1})$ contributes only to $\Lambda$ while
$a({\bf 10})$ does not contribute to the nucleon states.
Denoting by $N_1$, $N_2$, $N_3$, and $N_4$, the normalization constants for
the $(p,n)$, $(\Xi^0,\Xi^-)$, $(\Sigma^{\pm},\Sigma^0)$, and $\Lambda$
isospin multiplets, one has

\begin{mathletters}
\label{seis}
\begin{equation}
N^2_1 = N^2_0 + a^2({\bf\bar{10}}),
\ \ \ \ \ 
N^2_2 = N^2_0 + a^2({\bf 10}),
\end{equation}
\begin{equation}
N^2_3 = N^2_0 + \sum_{N={\bf 10,\bar{10}}}a^2(N),
\ \ \ \ \ 
N^2_4 = N^2_0 + a^2({\bf 1}),
\end{equation}

\noindent
where,

\begin{equation}
N^2_0 = 1 + \sum_{N={\bf 8_D,8_F,27}}a^2(N).
\end{equation}
\end{mathletters}

For example, using Tables~\ref{tabla1} and \ref{tabla2}, and Eq.~(\ref{seis}),
the physical spin-up proton state as given by Eq.~(\ref{dos}) is

\begin{eqnarray}
N_1 |p\uparrow\rangle 
&=&
|\tilde{p}\uparrow\rangle S({\bf 1}) 
+ \bar{\beta}_1 |\tilde{p}\uparrow\rangle S_{\eta}
+ \bar{\beta}_2 |\tilde{\Lambda}\uparrow\rangle S_{K^+}
\nonumber\\
&&
+ \bar{\beta}_3 |(\tilde{N}\uparrow S_{\pi})_{1/2,1/2}\rangle 
+ \bar{\beta}_4 |(\tilde{\Sigma}\uparrow S_K)_{1/2,1/2}\rangle.
\label{siete}
\end{eqnarray}

\noindent
Other baryon wavefunctions will have a similar structure.
Also, $(\tilde{N}\uparrow S_{\pi})_{1/2,1/2}$
($(\tilde{\Sigma}\uparrow S_{K})_{1/2,1/2}$) stand for the $I=I_3=1/2$
combination of the $I=1/2$ $\tilde{N}$ ($S_K$) and $I=1$ $S_{\pi}$
($\tilde{\Sigma}$) multiplets.
For any operator $\hat{O}$ which depends only on quarks, the matrix elements
are easily obtained using the ortogonality of the sea components.
Clearly $\langle p\uparrow|\hat{O}|p\uparrow\rangle $ will be a linear
combination of the matrix elements
$\langle \tilde{B}\uparrow|\hat{O}|\tilde{B'}\uparrow\rangle$ (known from
SQM) with coefficients which depend on the coefficients in the wavefunction.
Note, for the physical baryons to have $J^P=\frac{1}{2}^+$ the scalar
sea has $J^P=0^+$ since $\tilde{B}$ have $J^P=\frac{1}{2}^+$.

For applications, we need the quantities $(\Delta q)^B$, $q=u,d,s$; for each
spin-up baryon $B$.
These are defined as 

\begin{equation}
(\Delta q)^B =
n^B(q\uparrow)-n^B(q\downarrow)+n^B(\bar{q}\uparrow)-n^B(\bar{q}\downarrow),
\label{ocho}
\end{equation}

\noindent
where $n^B(q\uparrow)$ ($n^B(q\downarrow)$) are the number of spin-up
(spin-down) quarks of flavor $q$ in the spin-up baryon $B$.
Also, $n^B(\bar{q}\uparrow)$ and $n^B(\bar{q}\downarrow)$ have a similar
meaning for antiquarks.
However, these are zero as there are no explicit antiquarks in the
wavefunctions given by Eq.~(\ref{dos}).
The expressions for $(\Delta q)^B$ are given in Table~\ref{tabla3} in terms of
the coefficients $\bar{\beta}_i$, etc.
The expressions for $(\Delta q)^B$ reduce to the SQM values if there is no sea
contribution, that is, $a(N)=0$,
$N={\bf 1,8_F,8_D,10,\bar{10},27}$.
Moreover, the total spin $S_Z$ of a baryon is given by
$S^B_Z = (1/2) \sum_q (\Delta q)^B$.
For $S^B_Z=1/2$, we expect $\sum_q (\Delta q)^B = 1$ for a purely scalar sea.
This is indeed true for each baryon as can be seen from Table~\ref{tabla3}.
There are three $(\Delta q)^B$ $(q=u,d,s)$ for each $(Y,I)$-multiplet.
These twelve quantities and $(\Delta q)^{\Sigma^0\Lambda}$ are given in terms
of the six parameters of Eq.~(\ref{dos}) as our spin 1/2 baryons do not
belong to a definite representation of $SU(3)$.
To obtain a flavor octet physical baryon one restricts $N$ to ${\bf 8_F}$ and
${\bf 8_D}$ in Eq.~(\ref{dos}), that is, put $a(N)=0$ for
$N={\bf 27,10,\bar{10},1}$, so that the twelve $(\Delta q)^B$ are given in
terms of two parameters $a(N)$ with $N={\bf 8_F,8_D}$.
It is clear that our wavefunction, Eq.~(\ref{dos}), provides an explicit
model for spin 1/2 baryons to be compared to the phenomenological model
considered by some authors\cite{9} recently to fit the baryon magnetic moments.
These authors take the three quantities $(\Delta q)^p$ $(q=u,d,s)$ as
parameters to be determined from data but use flavor $SU(3)$ to express all
the other $(\Delta q)^B$ in terms of the $(\Delta q)^p$.
In our case the various $(\Delta q)^B$ are not simply related by flavor $SU(3)$
because of the non-trivial flavor properties of the sea and thus provides an
explicit and very different model for the baryons.

\section{Application to magnetic moments}

We assume the baryon magnetic moment operator $\hat{\mu}$ to be expressed
solely in terms of quarks as is usual in the quark model.
So that $\hat{\mu}=\sum_q (e_q/2m_q) \sigma^q_Z$ ($q=u,d,s$).
It is clear from Eq.~(\ref{dos}) that $\mu_B=\langle B|\hat{\mu}|B\rangle $
will be a linear combination of $\mu_{\tilde{B}}$ and
$\mu_{\tilde{\Sigma}^0\tilde{\Lambda}}$ weighted by the coefficients which
depend on $a(N)$'s.
The magnetic moments $\mu_{\tilde{B}}$ and the transition moment
$\mu_{\tilde{\Sigma}^0\tilde{\Lambda}}$ (for the core baryons) are given in
terms of the quark magnetic moments $\mu_q$ as per SQM.
For example,
$\mu_{\tilde{p}}=(4\mu_u-\mu_d)/3$,
$\mu_{\tilde{\Lambda}}=\mu_s$,
$\mu_{\tilde{\Sigma}^0\tilde{\Lambda}}=(\mu_u-\mu_d)/\sqrt{3}$,
etc.
Consequently, all the magnetic moments and the $\Sigma^0\rightarrow\Lambda$
transition magnetic moment in our model can be written simply as

\begin{mathletters}
\label{nueve}
\begin{eqnarray}
\mu_B &=& \sum_q (\Delta q)^B \mu_q,\ \ \ \ (q=u,d,s);
\\
\mu_{\Sigma^0\Lambda} &=&
\sum_q (\Delta q)^{\Sigma^0\Lambda} \mu_q,\ \ \ \ (q=u,d);
\end{eqnarray}
\end{mathletters}

\noindent
where the $(\Delta q)^B$ and $(\Delta q)^{\Sigma^0\Lambda}$ are given in
Table~\ref{tabla3} and $B=p,n,\Lambda,\dots$.

A class of models\cite{9} have been recently considered in which the magnetic
moments were expressed in terms of $\mu_q$ and $(\Delta q)^p$ ($q=u,d,s$)
without giving an explicit wavefunction.
Interestingly, Eq.~(\ref{nueve}) has the same general structure except that
here the twelve $(\Delta q)^B$ and $(\Delta q)^{\Sigma^0\Lambda}$ are not
related but depend on six parameters, namely, the six $a(N)$'s for the scalar
sea.
Despite, our general wavefunction, the isospin sum rule\cite{11}

\begin{equation}
\mu_{\Sigma^0}=\frac{1}{2}(\mu_{\Sigma^+}+\mu_{\Sigma^-})
\label{diez}
\end{equation}

\noindent
holds.
This is because $\hat{\mu}$ transforms as $(I=0) \oplus (I=1)$ in isospin
space and the wavefunction of Eq.~(\ref{dos}) respects isospin thus giving
$(\Delta q)^{\Sigma^0}=((\Delta q)^{\Sigma^+}+(\Delta q)^{\Sigma^-})/2$
(see Table~\ref{tabla3}).

Further, if we require the physical baryon states given by Eq.~(\ref{dos}) to
transform as an $SU(3)$ octet (i.e., put $a(N)=0$,
$N={\bf 1,10,\bar{10},27}$) then the seven $\mu_B$'s and
$\mu_{\Sigma^0\Lambda}$ depend non-linearly on two parameters
($a({\bf 8_F})$ and $a({\bf 8_D})$) of the wavefunction and the three $\mu_q$'s.
Even so, three sum rules, namely

\begin{mathletters}
\label{once}
\begin{eqnarray}
\mu_p - \mu_n &=&
(\mu_{\Sigma^+} - \mu_{\Sigma^-}) - (\mu_{\Xi^0} - \mu_{\Xi^-})
\\
((4.70589019\pm 5\times 10^{-7})\ \mu_N)
&&
((4.217\pm 0.031)\ \mu_N)
\end{eqnarray}
\end{mathletters}

\begin{mathletters}
\label{doce}
\begin{eqnarray}
-6\mu_{\Lambda} &=&
\mu_{\Sigma^+} + \mu_{\Sigma^-} - 2(\mu_p + \mu_n + \mu_{\Xi^0} + \mu_{\Xi^-})
\\
((3.678\pm 0.024)\ \mu_N)
&&
((3.340\pm 0.039)\ \mu_N)
\end{eqnarray}
\end{mathletters}

\begin{mathletters}
\label{docep}
\begin{eqnarray}
2\sqrt{3}\mu_{\Sigma^0\Lambda} &=&
2(\mu_p - \mu_n) - (\mu_{\Sigma^+} - \mu_{\Sigma^-})
\\
((5.577\pm 0.277)\ \mu_N)
&&
((5.794\pm 0.027)\ \mu_N)
\end{eqnarray}
\end{mathletters}

\noindent
emerge.
These have been noted earlier in the context of other models\cite{6,9}.
The values of the two sides taken from data\cite{10} are shown in parentheses.
The reason these sum rules hold despite the number of parameters is because
they are a consequence of flavor $SU(3)$ since baryons form a $SU(3)$ octet
and $\hat{\mu}$ transforms as ${\bf 1}\oplus{\bf 8}$.
However, as can be seen, the first two $SU(3)$ sum rules are not well satisfied
experimentally.
To avoid them one could modify the $SU(3)$ transformation properties of
$\hat{\mu}$ or the baryons.
A group-theoretic analysis with the most general $\hat{\mu}$ which would
contribute to the magnetic moments of an octet was done by Dothan\cite{12}
over a decade ago.
Such a $\hat{\mu}$ could arise from $SU(3)$ breaking effects.
However, several authors\cite{13} have considered models in which they modify
the baryon wavefunction.
In our approach, we keep $\hat{\mu}$ as given by the quark model but modify
the baryon wavefunction by taking a sea with flavor and spin into account.

As noted above the poorly satisfied $SU(3)$ sum rules
Eqs.~(\ref{once}) and (\ref{doce}) will hold as long the physical baryon is
restricted to be an octet.
This means we must include $SU(3)$ breaking effects in the baryon wavefunction
by considering non-zero $a(N)$ with $N={\bf 1,10,\bar{10},27}$.

In making our fits we have used {\em experimental errors }as given by Particle
Data Group\cite{10}.
This is in contrast to many authors who use \lq\lq theoretical errors" of the
order of a few percent or more to fit the data.
In actual fact the experimental errors are very much smaller.
Furthermore, we keep in mind that the constituent quark masses are
$m_u,\ m_d\approx 300{\rm MeV}$, and $m_s\approx 500{\rm Mev}$, so we expect
$\mu_u\cong -2\mu_d> 0$ and $\mu_s\cong 0.6 \mu_d$.
Current quark masses would give a very different numerical range for the
ratios $\mu_u/\mu_d$ and $\mu_s/\mu_d$.
Also, if the core baryon contribution is dominant then the parameters
determining the sea should be small compared to unity.
Furthermore, for a dominantly $SU(3)$ octet physical baryon
$a({\bf 8_F})$ and $a({\bf 8_D})$ should be larger than the other parameters
in the wave function.

SQM has three parameters $\mu_q$ ($q=u,\ d,\ s$) the quark magnetic moments in
nuclear magnetons $\mu_N$.
A fit using experimental errors gives $\chi^2/\mbox{DOF}=1818/5$ with
$\mu_u=1.852$, $\mu_d=-0.972$, $\mu_s=-0.701$.

The situation improves a little for a pure octet physical baryon with scalar
sea described by $a({\bf 8_F})$ and $a({\bf 8_D})$.
These two sea parameters enter Eqs.~(\ref{nueve}) only through the three
combinations given by $(\Delta q)^p$.
Hence, the 3 sum rules in Eqs.~(\ref{once})-(\ref{docep}).
For {\em experimental errors }with $\mu_q$ also as parameters one
obtains $\chi^2/\mbox{DOF}=652/3$.
Most of the contribution to $\chi^2$ comes from a poor fit to $\mu_{\Sigma^+}$,
$\mu_{\Sigma^-}$, and $\mu_{\Xi^0}$.
This is a clear indication that admixture of other $SU(3)$ representations in
our wavefunction need to be considered.

To get a feeling for how the sea contributes we did extensive and systematic
numerical analysis.
In all the fits, in addition to the sea parameters, $\mu_q$ were treated as
parameters.

In general, there are six parameters $a(N)$'s,
($N={\bf 1}$, ${\bf 8_F}$, ${\bf 8_D}$, ${\bf 10}$, ${\bf \bar{10}}$,
${\bf 27}$), in the wavefunction and the three $\mu_q$'s.
These nine parameters provide a perfect fit with $\chi^2=1.5\times 10^{-4}$.
This clearly means that the scalar sea contribution modifies the values of
$(\Delta q)^B$ in the right direction for a fit to the baryon magnetic moments.
However, a seven parameter fit with
$a({\bf 1})=0.0625$,
$a({\bf 8_D})=-0.1558$,
$a({\bf 8_F})=0.1896$,
$a({\bf 10})=0.4297$,
and
$\mu_u=1.85893$,
$\mu_d=-0.99884$,
and
$\mu_s=-0.65302$,
provides an excellent fit with $\chi^2/\mbox{DOF}=0.838$.
The $\mu_q$ (in units of $\mu_N$) imply for the quark masses the values
$m_u=336.49\ {\rm MeV}$,
$m_d=313.12\ {\rm MeV}$,
and
$m_s=478.94\ {\rm MeV}$,
which are in accord with the constituent quark model.
A noteworthy six parameter fit with $\chi^2/\mbox{DOF}=5.60/2$ is given by
$a({\bf 8_D})=-0.2262$,
$a({\bf 8_F})=0.2776$,
and
$a({\bf 10})=0.4216$,
with
$\mu_u=1.86688$,
$\mu_d=-1.02558$,
and
$\mu_s=-0.64663$.
The predictions of this six parameter fit are displayed in the
\lq\lq Scalar sea" column of Table~\ref{tabla4}.
This fit gives
$(\Delta u)^p=1.322$,
$(\Delta d)^p=-0.304$,
and
$(\Delta s)^p=-0.018$,
which are near SQM values with a small strangeness content in the nucleon.

\section{Summary}

In summary, we have considered the physical spin 1/2 low-lying baryons to be
formed out \lq\lq core" baryons (described by the $q^3$-wavefunction of SQM)
and a flavor octet but color singlet \lq\lq sea".

This sea (which may contain arbitrary number of gluons and $q\bar{q}$-pairs)
is specified only by its total flavor and spin quantum numbers.
The most general wavefunction for the physical baryons for an octet sea with
spin 0 was considered which respected isospin and
hypercharge (or strangeness).
Owing to the flavor properties of the sea the nucleons can have a non-zero
strange quark content (giving $(\Delta s)^p=(\Delta s)^n\neq 0$) through the
strange core baryons.
In this model the eight baryons no longer form an exact $SU(3)$ octet.
The admixture of other flavor $SU(3)$ representations in the wavefunction is
understood to represent broken $SU(3)$ effects.
The parameters in the wavefunction describing the sea were determined by
application to the available data on baryon magnetic moments.
To these 8 pieces of data we found good fits with six parameters
{\em using available experimental errors}~\cite{10}.
Three of these parameters determined the sea contribution while the other three
were $\mu_q$'s (or $m_q$'s, $q=u, d, s$) the quark magnetic moments (masses).
The modified baryon wavefunction including such a sea suggested by the fits to
the data is simply

\begin{equation}
B(1/2\uparrow)
=\tilde{B}({\bf 8},1/2\uparrow)H_{0}S({\bf 1})
+\sum_{N={\bf 8_{D},8_{F},10}}a(N)\left[\tilde{B}({\bf 8},1/2\uparrow)H_{0}
\otimes S({\bf 8})\right]_{N}
\label{diecisiete}
\end{equation}

\noindent
Our results suggest that the physical spin 1/2 \lq\lq octet" baryons contain
an admixture of primarily the {\bf 10} representation.
Why $SU(3)$ breaking (which we have invoked through a flavor octet sea)
induce this representation is a question for the future when one is able to
calculate the parameters in the wavefunction of Eq.~(\ref{dos}) reliably from
quantum chromodynamics.

We thank CONACyT (M\'exico) for partial support.

\mediumtext
\begin{table}
\caption{Contribution to the physical baryon state $B(Y,I,I_3)$ formed out of
$\tilde{B}(Y,I,I_3)$ and flavor octet states $S(Y,I,I_3)$ (see second term
in Eq.~(2)).
The core baryon states $\tilde{B}$ denoted by $\tilde{p}$, $\tilde{n}$, etc.
are the normal 3 valence quark states of SQM.
The sea octet states are denoted by $S_{\pi^+}=S(0,1,1)$, etc.\ as in Eq.~(3).
Further, $(\tilde{N}S_{\pi})_{I,I_3}$, $(\tilde{\Sigma}S_{\bar{K}})_{I,I_3}$,
$(\tilde{\Sigma}S_{\pi})_{I,I_3}$, $\dots$ stand for total $I$, $I_3$
{\em normalized} combinations of $\tilde{N}$ and $S_{\pi}$, etc.
See Table~II for the coefficients $\bar{\beta}_i$, $\beta_i$, $\gamma_i$,
and $\delta_i$.}
\label{tabla1}
\begin{tabular}
{
cc
}
$B(Y,I,I_3)$ &
$\tilde{B}(Y,I,I_3)$ and $S(Y,I,I_3)$
\\
\hline
\\
$p$ &
$
\bar{\beta}_1 \tilde{p} S_{\eta} 
+ \bar{\beta}_2 \tilde{\Lambda} S_{K^+} 
+ \bar{\beta}_3 (\tilde{N}S_{\pi})_{1/2,1/2}
+ \bar{\beta}_4 (\tilde{\Sigma}S_{K})_{1/2,1/2}
$
\\\\
$n$ &
$
\bar{\beta}_1 \tilde{n} S_{\eta} 
+ \bar{\beta}_2 \tilde{\Lambda} S_{K^0} 
+ \bar{\beta}_3 (\tilde{N}S_{\pi})_{1/2,-1/2}
+ \bar{\beta}_4 (\tilde{\Sigma}S_{K})_{1/2,-1/2}
$
\\\\
$\Xi^0$ &
$
\beta_1 \tilde{\Xi}^0 S_{\eta} 
+ \beta_2 \tilde{\Lambda} S_{\bar{K}^0} 
+ \beta_3 (\tilde{\Xi}S_{\pi})_{1/2,1/2}
+ \beta_4 (\tilde{\Sigma}S_{\bar{K}})_{1/2,1/2}
$
\\\\
$\Xi^-$ &
$
\beta_1 \tilde{\Xi}^- S_{\eta} 
+ \beta_2 \tilde{\Lambda} S_{\bar{K}^-} 
+ \beta_3 (\tilde{\Xi}S_{\pi})_{1/2,-1/2}
+ \beta_4 (\tilde{\Sigma}S_{\bar{K}})_{1/2,-1/2}
$
\\\\
$\Sigma^+$ &
$
\gamma_1 \tilde{p} S_{\bar{K}^0} 
+ \gamma_2 \tilde{\Xi}^0 S_{K^+} 
+ \gamma_3 \tilde{\Lambda} S_{\pi^+}
+ \gamma_4 \tilde{\Sigma}^+ S_{\eta}
+ \gamma_5 (\tilde{\Sigma}S_{\pi})_{1,1}
$
\\\\
$\Sigma^-$ &
$
\gamma_1 \tilde{n} S_{K^-} 
+ \gamma_2 \tilde{\Xi}^- S_{K^0} 
+ \gamma_3 \tilde{\Lambda} S_{\pi^-}
+ \gamma_4 \tilde{\Sigma}^- S_{\eta}
+ \gamma_5 (\tilde{\Sigma}S_{\pi})_{1,-1}
$
\\\\
$\Sigma^0$ &
$
\gamma_1 (\tilde{N} S_{\bar{K}})_{1,0} 
+ \gamma_2 (\tilde{\Xi} S_{K})_{1,0} 
+ \gamma_3 \tilde{\Lambda} S_{\pi^0}
+ \gamma_4 \tilde{\Sigma}^0 S_{\eta}
+ \gamma_5 (\tilde{\Sigma}S_{\pi})_{1,0}
$
\\\\
$\Lambda$ &
$
\delta_1 (\tilde{N} S_{\bar{K}})_{0,0} 
+ \delta_2 (\tilde{\Xi} S_{K})_{0,0} 
+ \delta_3 \tilde{\Lambda} S_{\eta}
+ \delta_4 (\tilde{\Sigma}S_{\pi})_{0,0}
$
\\\\
\end{tabular}
\end{table}

\mediumtext
\begin{table}
\squeezetable
\caption{The coefficients $\bar{\beta}_i$, $\beta_i$, $\gamma_i$, and
$\delta_i$ in Table~I expressed in terms of the coefficients $a(N)$,
$N={\bf 1}$, ${\bf 8_{F}}$, ${\bf 8_{D}}$, ${\bf 10}$, ${\bf \bar{10}}$,
${\bf 27}$, in the $2^{\rm nd}$ term (from scalar sea) in Eq.~(2).}
\label{tabla2}
\begin{tabular}
{
c
}
$
\beta_1=\frac{1}{\sqrt {20}}(3a({\bf 27})-a({\bf 8_D}))-
\frac{1}{2}(a({\bf 8_F})-a({\bf 10}))
$
\\\\
$
\beta_2=\frac{1}{\sqrt {20}}(3a({\bf 27})-a({\bf 8_D}))+
\frac{1}{2}(a({\bf 8_F})-a({\bf 10}))
$
\\\\
$
\beta_3=-\frac{1}{\sqrt {20}}(a({\bf 27})+3a({\bf 8_D}))+
\frac{1}{2}(a({\bf 8_F})+a({\bf 10}))
$
\\\\
$
\beta_4=\frac{1}{\sqrt{20}}(a({\bf 27})+3a({\bf 8_D}))+
\frac{1}{2}(a({\bf 8_F})+a({\bf 10}))
$
\\\\
$
\bar{\beta}_1=\frac{1}{\sqrt {20}}(3a({\bf 27})-a({\bf 8_D}))+
\frac{1}{2}(a({\bf 8_F})+a({\bf \bar{10}}))
$
\\\\
$
\bar{\beta}_2=\frac{1}{\sqrt {20}}(3a({\bf 27})-a({\bf 8_D}))-
\frac{1}{2}(a({\bf 8_F})+a({\bf \bar{10}}))
$
\\\\
$
\bar{\beta}_3=\frac{1}{\sqrt {20}}(a({\bf 27})+3a({\bf 8_D}))+
\frac{1}{2}(a({\bf 8_F})-a({\bf \bar{10}}))
$
\\\\
$
\bar{\beta}_4 =-\frac{1}{\sqrt {20}}(a({\bf 27})+3a({\bf 8_D}))+
\frac{1}{2}(a({\bf 8_F})-a({\bf \bar{10}}))
$
\\\\
$
\gamma_1=\frac{1}{\sqrt{10}}(\sqrt{2}a({\bf 27})-\sqrt{3}a({\bf 8_D}))+
\frac{1}{\sqrt{6}}(a({\bf 8_F})-a({\bf 10})+a({\bf\bar{10}}))
$
\\\\
$
\gamma_2=\frac{1}{\sqrt{10}}(\sqrt{2}a({\bf 27})-\sqrt{3}a({\bf 8_D}))-
\frac{1}{\sqrt{6}}(a({\bf 8_F})-a({\bf 10})+a({\bf\bar{10}}))
$
\\\\
$
\gamma_3=\frac{1}{\sqrt{10}}(\sqrt{3}a({\bf 27})+\sqrt{2}a({\bf 8_D}))-
\frac{1}{2}(a({\bf 10})+a({\bf\bar{10}}))
$
\\\\
$
\gamma_4=\frac{1}{\sqrt{10}}(\sqrt{3}a({\bf 27})+\sqrt{2}a({\bf 8_D}))+
\frac{1}{2}(a({\bf 10})+a({\bf\bar{10}}))
$
\\\\
$
\gamma_5=\frac{1}{\sqrt{6}}(2a({\bf 8_F})+a({\bf 10})-a({\bf\bar{10}}))
$
\\\\
$
\delta_1=\frac{1}{\sqrt{20}}(\sqrt{3}a({\bf 27})+\sqrt{2}a({\bf 8_D}))+
\frac{1}{2}(\sqrt{2}a({\bf 8_F})+a({\bf 1}))
$
\\\\
$
\delta_2=-\frac{1}{\sqrt{20}}(\sqrt{3}a({\bf 27})+\sqrt{2}a({\bf 8_D}))+
\frac{1}{2}(\sqrt{2}a({\bf 8_F})-a({\bf 1}))
$
\\\\
$
\delta_3=\frac{3\sqrt{3}}{\sqrt{40}}a({\bf 27})-
\frac{1}{\sqrt{5}}a({\bf 8_D})-
\frac{\sqrt{2}}{4}a({\bf 1})
$
\\\\
$
\delta_4=-\frac{1}{\sqrt{40}}a({\bf 27})-\sqrt{\frac{3}{5}}a({\bf 8_D})+
\frac{\sqrt{6}}{4}a({\bf 1})
$
\\\\
\end{tabular}
\end{table}

\mediumtext
\begin{table}
\squeezetable
\caption{$(\Delta q)^B$ defined in Eq.~(8) for physical baryon $B$ given by
general wavefunction in Eq.~(2).
The normalizations $N_1$, $N_2$, $N_3$, and $N_4$ are given in Eqs.~(6).
The $(\Delta q)^{\Sigma^0\Lambda}$ for the $\Sigma^0\rightarrow\Lambda$
transition magnetic moment is also given.}
\label{tabla3}
\begin{tabular}
{
c
}
$
(\Delta u)^p=
\frac{1}{3N^2_1}
(
4
+
4\bar{\beta}^2_1+\frac{2}{3}\bar{\beta}^2_3+\frac{10}{3}\bar{\beta}^2_4
-2\bar{\beta}_2\bar{\beta}_4
)
$
\\\\
$
(\Delta d)^p=
\frac{1}{3N^2_1}
(
-1
-
\bar{\beta}^2_1+\frac{7}{3}\bar{\beta}^2_3+\frac{2}{3}\bar{\beta}^2_4
+2\bar{\beta}_2\bar{\beta}_4
)
\ \ \ \ \ \ 
(\Delta s)^p=
\frac{1}{3N^2_1}
(
3\bar{\beta}^2_2-\bar{\beta}^2_4
)
$
\\\\
$
(\Delta u)^n=(\Delta d)^p
\ \ \ \ \ \ 
(\Delta d)^n=(\Delta u)^p
\ \ \ \ \ \ 
(\Delta s)^n=(\Delta s)^p
$
\\\\
$
(\Delta u)^{\Xi^0}=
\frac{1}{3N^2_2}
(
-1
-
\beta^2_1-\frac{1}{3}\beta^2_3+\frac{10}{3}\beta^2_4
-2\beta_2\beta_4
)
$
\\\\
$
(\Delta d)^{\Xi^0}=
\frac{1}{3N^2_2}
(
-\frac{2}{3}\beta^2_3+\frac{2}{3}\beta^2_4
+2\beta_2\beta_4
)
\ \ \ \ \ \ 
(\Delta s)^{\Xi^0}=
\frac{1}{3N^2_2}
(
4
+
4\beta^2_1+3\beta^2_2+4\beta^2_3-\beta^2_4
)
$
\\\\
$
(\Delta u)^{\Xi^-}=(\Delta d)^{\Xi^0}
\ \ \ \ \ \ 
(\Delta d)^{\Xi^-}=(\Delta u)^{\Xi^0}
\ \ \ \ \ \ 
(\Delta s)^{\Xi^-}=(\Delta s)^{\Xi^0}
$
\\\\
$
(\Delta u)^{\Sigma^+}=
\frac{1}{3N^2_3}
(
4
+
4\gamma^2_1-\gamma^2_2+4\gamma^2_4+3\gamma^2_5
-\sqrt{6}\gamma_3\gamma_5
)
$
\\\\
$
(\Delta d)^{\Sigma^+}=
\frac{1}{3N^2_3}
(
-\gamma^2_1+\gamma^2_5
+\sqrt{6}\gamma_3\gamma_5
)
\ \ \ \ \ \ 
(\Delta s)^{\Sigma^+}=
\frac{1}{3N^2_3}
(
-1
+
4\gamma^2_2+3\gamma^2_3-\gamma^2_4-\gamma^2_5
)
$
\\\\
$
(\Delta u)^{\Sigma^-}=(\Delta d)^{\Sigma^+}
\ \ \ \ \ \ 
(\Delta d)^{\Sigma^-}=(\Delta u)^{\Sigma^+}
\ \ \ \ \ \ 
(\Delta s)^{\Sigma^-}=(\Delta s)^{\Sigma^+}
$
\\\\
$
(\Delta u)^{\Sigma^0}=
\frac{1}{2}[(\Delta u)^{\Sigma^+}+(\Delta u)^{\Sigma^-}]
\ \ \ \ \ \ 
(\Delta d)^{\Sigma^0}=(\Delta u)^{\Sigma^0}
\ \ \ \ \ \ 
(\Delta s)^{\Sigma^0}=(\Delta s)^{\Sigma^+}
$
\\\\
$
(\Delta u)^{\Lambda}=
\frac{1}{3N^2_4}
(
\frac{3}{2}\delta^2_1-\frac{1}{2}\delta^2_2+2\delta^2_4
)
$
\\\\
$
(\Delta d)^{\Lambda}=(\Delta u)^{\Lambda}
\ \ \ \ \ \ \ \ \ 
(\Delta s)^{\Lambda}=
\frac{1}{3N^2_4}
(
3
+
4\delta^2_2+3\delta^2_3-\delta^2_4
)
$
\\\\
$
(\Delta u)^{\Sigma^0\Lambda}=
\frac{1}{N_3N_4}
(
\frac{1}{\sqrt{3}}
+
\frac{1}{\sqrt{3}}\gamma_4\delta_3-\frac{1}{3}\gamma_3\delta_4
+
\frac{5}{6}\gamma_1\delta_1-\frac{1}{6}\gamma_2\delta_2
+\frac{4}{3\sqrt{6}}\gamma_5\delta_4
)
$
\\\\
$
(\Delta d)^{\Sigma^0\Lambda}=-(\Delta u)^{\Sigma^0\Lambda}
\ \ \ \ \ \ \ \ \ \ \ \ 
(\Delta s)^{\Sigma^0\Lambda}=0
$
\\\\
\end{tabular}
\end{table}

\mediumtext
\begin{table}
\caption{Predictions for the baryon magnetic moments (in N.M.), along with the
eight experimental measurements currently available.}
\label{tabla4}
\begin{tabular}
{
c
r@{\,$\pm$\,}l
d
d
}
Baryon &
\multicolumn{2}{c}{Data} &
SQM &
Scalar sea
\\
\hline
\\
$p$ &
2.79284739 & $6\times 10^{-8}$ &
2.7928 &
2.7928 
\\\\
$n$ &
$-$1.9130428 & $5\times 10^{-7}$ &
$-$1.9130 &
$-$1.9130 
\\\\
$\Lambda$ &
$-$0.613 & 0.004 &
$-$0.701 &
$-$0.616 
\\\\
$\Sigma^+$ &
2.458 & 0.010 &
2.703 &
2.456 
\\\\
$\Sigma^0$ &
\multicolumn{2}{c}{--------} &
0.8203 &
0.6238 
\\\\
$\Sigma^-$ &
$-$1.160 & 0.025 &
$-$1.062 &
$-$1.208 
\\\\
$\Xi^0$ &
$-$1.250 & 0.014 &
$-$1.552 &
$-$1.248 
\\\\
$\Xi^-$ &
$-$0.6507 & 0.0025 &
$-$0.6111 &
$-$0.6500 
\\\\
$|\Sigma^0\rightarrow\Lambda|$ &
1.61 & 0.08 &
1.63 &
1.52 
\\\\
$\chi^2/\mbox{DOF}$ &
\multicolumn{2}{c}{--------} &
1818/5 &
5.60/2 
\\\\
\end{tabular}
\end{table}


\begin{references}
\bibitem{7}
This is in contrast to Ref.\cite{6} where a flavorless sea with total spin and
color quantum numbers was considered.
\end{references}
\end{document}